\documentclass[prd, twocolumn, superscriptaddress, nofootinbib, 
floatfix]{revtex4-2}

\usepackage{lipsum}
\usepackage{xcolor}
\usepackage{etoolbox}
\usepackage{graphicx}
\usepackage{wrapfig}
\usepackage{amsmath}
\usepackage{amsfonts}
\usepackage{amssymb}
\usepackage{multirow}
\usepackage{slashed}
\usepackage{physics}
\usepackage{ulem}
\usepackage{diagbox}
\usepackage{float}

\usepackage[colorlinks=true, linkcolor=blue, citecolor=blue, urlcolor=blue]{hyperref}

\DeclareRobustCommand{\Eq}[1]{Eq.~\eqref{eq:#1}}

\DeclareRobustCommand{\fig}[1]{Fig.~\ref{fig:#1}}

\DeclareRobustCommand{\app}[1]{App.~\ref{app:#1}}

\DeclareRobustCommand{\tb}[1]{Table~\ref{tb:#1}}

\DeclareRobustCommand{\refcite}[1]{Ref.~\cite{#1}}

\newcommand\bets{\begin{table*}}
\newcommand\eets[1]{\label{tb:#1}\end{table*}}
\begin{document}

\preprint{APS/123-QED}

\title{Systematic Uncertainties from Gribov Copies in Lattice Calculation of Parton Distributions in the Coulomb gauge}

\author{Xiang Gao}
\affiliation{Physics Division, Argonne National Laboratory, Lemont, IL 60439, USA}

\author{Jinchen He}
\email{jinchen@umd.edu}
\affiliation{Maryland Center for Fundamental Physics, University of Maryland, College Park, MD 20742, USA}
\affiliation{Physics Division, Argonne National Laboratory, Lemont, IL 60439, USA}

\author{Rui Zhang}
\affiliation{Physics Division, Argonne National Laboratory, Lemont, IL 60439, USA}

\author{Yong Zhao}
\affiliation{Physics Division, Argonne National Laboratory, Lemont, IL 60439, USA}

\date{\today}

\begin{abstract}
Recently, it has been proposed to compute parton distributions from boosted correlators fixed in the Coulomb gauge within the framework of Large-Momentum Effective Theory. This method does not involve Wilson lines and could greatly improve the efficiency and precision of lattice QCD calculations. However, there are concerns about whether the systematic uncertainties from Gribov copies, which correspond to the ambiguity in lattice gauge-fixing, are under control. This work gives an assessment of the Gribov copies' effect in the Coulomb-gauge-fixed quark correlators. We utilize different strategies for the Coulomb-gauge fixing, selecting two different groups of Gribov copies based on the lattice gauge configurations. We test the difference in the resulted spatial quark correlators in the vacuum and a pion state. Our findings indicate that the statistical errors of the matrix elements from both Gribov copies, regardless of the correlation range, decrease proportionally to the square root of the number of gauge configurations. The difference between the strategies does not show statistical significance compared to the gauge noise. This demonstrates that the effect of the Gribov copies can be neglected in the practical lattice calculation of the quark parton distributions.
\end{abstract}

\maketitle

\section{Motivation}
\label{sec:motivation}

Understanding the 3D partonic structure of the proton is among the top goals of the experiments at RHIC~\cite{RHICSPIN:2023zxx}, Jefferson Lab~\cite{Dudek:2012vr}, Fermilab~\cite{SeaQuest:2017kjt,SeaQuest:2019hsx}, CERN~\cite{COMPASS:2007rjf} and the future Electron-Ion Collider~\cite{Accardi:2012qut,AbdulKhalek:2021gbh}.
In recent years, significant progress has been made towards 3D tomography of hadrons from lattice quantum chromodynamics (QCD)~\cite{Constantinou:2020hdm}. Among various approaches, Large-Momentum Effective Theory (LaMET)~\cite{Ji:2013dva,Ji:2014gla,Ji:2020ect} provides a first-principles framework to calculate the $x$-dependence of all parton distributions, which has led to substantial lattice results on the parton distribution functions (PDFs), generalized parton distributions (GPDs) and transverse-momentum-dependent distributions (TMDs). (See, for example, Refs.~\cite{Gao:2021dbh,LatticeParton:2022xsd,Bhattacharya:2022aob,Holligan:2023jqh,LatticePartonCollaborationLPC:2022myp,Avkhadiev:2024mgd}.)

In the LaMET framework, one approaches the light-cone parton distribution through a quasi-distribution defined by an equal-time correlator of quarks or gluons in a boosted hadron state, which is usually contracted with a Wilson line that makes it gauge invariant. In the TMD case, the Wilson line consists of two longitudinal gauge links with long extension and a transverse link that closes the contour to ensure gauge invariance. For long-range correlations, the self-energy of Wilson lines imposes an exponential suppression of the signal-to-noise ratio in lattice simulations due to the linear power divergences~\cite{Ji:2017oey,Ishikawa:2017faj,Green:2017xeu}, which becomes a limitation on the precision of computations, especially for TMDs in the non-perturbative region.

Recently, a new method was proposed to compute parton distributions from correlators fixed in the Coulomb gauge (CG) without Wilson lines~\cite{Gao:2023lny,Zhao:2023ptv,Bollweg:2024zet}. These boosted CG correlators belong to the same universality class~\cite{Hatta:2013gta,Ji:2020ect} as the gauge-invariant quasi-distributions and can, therefore, be matched to light-cone TMDs through LaMET~\cite{Ebert:2019okf,Ji:2019sxk,Ji:2019ewn,Ebert:2022fmh,Zhao:2023ptv}. The CG method can significantly alleviate the suppression of the signal-to-noise ratio in long-range correlations, thus dramatically improving the precision of the TMD calculation. Moreover, lattice renormalization of the correlators is also greatly simplified with the absence of linear divergence and the reduction of operator mixings~\cite{Gao:2023lny}, and the 3D rotational symmetry of the CG condition allows one to access the same hadron momentum from an off-axis direction with less computational cost~\cite{Gao:2023lny}.
Nevertheless, there are concerns on whether the CG fixing process introduces extra systematic uncertainties that cannot be controlled. The main uncertainty comes from the effect of the so-called Gribov copies~\cite{Gribov:1977wm,Singer:1978dk}, which are multiple solutions on a gauge orbit that satisfies the CG condition. Since the standard lattice gauge-fixing algorithm~\cite{Mandula:1987rh,Davies:1987vs} does not uniquely fix the CG, it is natural to ask if the variation of the gauge-dependent matrix elements from different Gribov copies is significant. If such a variation is much smaller than the statistical errors from averaging over the set of gauge configurations in lattice simulation, then the Gribov copies do not pose a practical problem for their application to parton physics. This work aims at assessing such a systematic uncertainty from Gribov copies for the CG quark correlators.

In lattice QCD calculations, the path integral in the continuum theory is replaced by an ordinary integral over the gauge link variables $U_\mu(x)$ on a finite lattice, so gauge-invariant quantities can be solved non-perturbatively without gauge-fixing. Gauge-fixing is only required when calculating gauge-dependent quantities, such as the quark propagators. To fix the gauge on the lattice, the standard method is to take the extreme point of a specific functional~\cite{Mandula:1987rh,Davies:1987vs}. Take the CG as an example. In continuum theory, one can define a functional as follows:
\begin{align}
    F_{\text{CG}}[A, \Omega] \equiv \frac{1}{2} \sum_{\mu = 1}^{3} \int d^4 x A^a_{\Omega \mu} (x) A_{\Omega}^{\mu a} (x) ~,
\end{align}
where $a$ is the SU(3) color index in the adjoint representation, $A_{\Omega \mu} (x) \equiv \Omega^\dagger (x) A_\mu(x) \Omega(x) + \frac{i}{g} \Omega^\dagger (x) \partial_\mu \Omega(x)$, and $\Omega(x)$ is a gauge transformation. If an infinitesimal gauge transformation is applied $\theta$ to the gauge potential $A_{\Omega \mu}$, 
\begin{align}
    \delta A_{\Omega \mu} &= - D^{\Omega}_{\mu a b} \theta_{b} = - (\partial_\mu \theta_a - g f^{c a b} A_{\Omega \mu}^c \theta_b ) ~,
\end{align}
then the variation of the functional is
\begin{align}
\begin{aligned}
    \delta F_{\text{CG}}[A, \Omega] 
    &= - \sum_{\mu = 1}^{3} \int d^4 x (\partial_\mu \theta_a - g f^{c a b} A_{\Omega \mu}^c \theta_b ) A_{\Omega}^{\mu a} \\
    &= \sum_{\mu = 1}^{3} \int d^4 x \theta_a (\partial_\mu A^{\mu a}_\Omega ) ~.
\end{aligned}
\end{align}
Therefore, if the functional reaches an extreme point, then the CG condition $\sum_{\mu = 1}^{3} \partial_\mu A^{\mu a}_\Omega = 0$ is satisfied. On the lattice, the discretized version of the functional is~\cite{Mandula:1987rh}
\begin{align}
    F[U, \Omega] \equiv - \Re\left[ \mathrm{Tr} \sum_{x} \sum_{\mu = 1}^{l}  \Omega^\dagger (x+\hat{\mu}) U_\mu (x) \Omega(x) \right] ~,
    \label{eq:functional}
\end{align}
where $l=4$ for Landau gauge and $l=3$ for CG.

However, as shown by Gribov~\cite{Gribov:1977wm} and Singer~\cite{Singer:1978dk}, the extrema of the functional are not unique in general, which are named Gribov copies. Generally, the impact of Gribov copies on the lattice calculation can be put into two categories: measurement distortion and lattice Gribov noise; the latter is not separable from the statistical uncertainty of the Monte Carlo simulation. An illustrative example of a distortion caused by the presence of Gribov copies can be observed in the evaluation of the photon propagator in compact U(1) within the Coulomb phase~\cite{Petrarca:1999cx}. The impact of Gribov copies has been discussed in the literature~\cite{Bali:1996dm,Giusti:2001xf,Maas:2008ri,Burgio:2016nad,Golterman:2012dx}, where it has been shown that the gluon propagator in the far infrared region is influenced by Gribov copies in the SU(2) gauge theory~\cite{Maas:2008ri,Bornyakov:2011fn}. Interestingly, in the quark propagator sector, the Gribov noise was found to be negligible compared to the statistical noise~\cite{Burgio:2012ph,Kalusche:2024osk}.

The paper is organized as follows. Our methodology for generating different Gribov copies and assessing their systematic uncertainties is explained in Sec.~\ref{sec:method}. The numerical results of the correlator for pion valence quark quasi-distributions and the light-quark spatial propagator are shown in Sec.~\ref{sec:numerical_results}. Finally, in Sec.~\ref{sec:conclusions}, we conclude that the systematic uncertainties from Gribov copies are negligible compared to the statistical noise from averaging over the gauge configurations, thus solidifying the CG correlator approach for calculating parton physics.

\section{Methodology}
\label{sec:method}

As mentioned in Sec.~\ref{sec:motivation}, different Gribov copies correspond to different extrema of the functional, i.e., different gauge transformations. Thus, a straightforward way to distinguish different Gribov copies is to compare the gauge transformations for gauge-fixing. Nevertheless, as a large vector, gauge transformations are not a good criterion in numerical processing. A better criterion is the functional value after gauge-fixing. Variations beyond the accuracy of the gauge-fixing imply a distinct extremum, which is a new Gribov copy. There are two possible criteria for the precision of the gauge-fixing, one is the normalized variation of the functional $\delta F / F$; the other is the residual gradient of the functional defined as~\cite{Giusti:2001xf}
\begin{align}
    \theta^G \equiv \frac{1}{V} \sum_x \theta^G(x) \equiv \frac{1}{V} \sum_x \operatorname{Tr}\left[\Delta^G(x)\left(\Delta^G\right)^{\dagger}(x)\right] ~,
\end{align}
where $V$ is the lattice volume and the gradient is defined as $\Delta^G(x) \equiv \sum_\mu\left(A_\mu^G(x)-A_\mu^G(x-\hat{\mu})\right)$. Both of them approach zero under the gauge-fixing condition. In this work, we take the first criterion such that we can distinguish different copies according to the value of $F$ after gauge-fixing.

Various techniques have been discussed in the literature for generating different Gribov copies~\cite{Giusti:2001xf}. One such method, known as the mother-and-daughter approach, involves applying $n$ distinct random gauge transformations to the mother configuration $M$, followed by fixing these transformed configurations to a particular gauge in order to produce $n$ daughter configurations $\{M_1,M_2,\ldots, M_n\}$. Typically, these daughters have the potential to access a maximum of $n$ distinct Gribov copies, with the possibility that multiple daughters may correspond to the same Gribov copy. An alternative approach involves using the overrelaxation algorithm~\cite{Mandula:1990vs}. Different choices of the overrelaxation parameter will lead to different Gribov copies even if one starts from the same initial configuration~\cite{Giusti:2001xf}. In this work, we adopt the mother-and-daughter method to generate different Gribov copies for each initial configuration. The same method has also been used in high-precision lattice calculations of nucleon axial coupling ~\cite{Chang:2018uxx}.

The numerical effects introduced by Gribov copies can be categorized into two primary types: lattice Gribov noise and measurement distortion~\cite{Giusti:2001xf}. Lattice Gribov noise arises from residual gauge freedoms and is related to the distribution of the target quantity across all Gribov copies. This noise is indistinguishable from the statistical uncertainties that arise from variations between lattice configurations~\cite{Martinelli:1993dq, Paciello:1994gs}. Consequently, the predominant concern regarding Gribov copies is the resultant measurement distortion. The distortion is related to the choice of the representative within various Gribov copies in gauge-fixing, i.e., the gauge-fixing strategy. A biased strategy can result in deviations from the true value, even though the Gribov noise remains indistinguishable from the statistical uncertainties. For example, selecting the copy with the maximal value of the target quantity as the representative for each configuration will clearly produce a distorted measurement. Therefore, it is necessary to quantify measurement distortion, for which a standard approach within the literature is to take various gauge-fixing strategies for comparative analysis. The resultant discrepancies among these strategies give an estimation of the measurement distortion.
As an illustration, Ref.~\cite{Maas:2008ri} examines two distinct strategies: {\bf Absolute gauge-fixing} means identifying the global minimum of the functional across all fixed copies; {\bf Minimal gauge-fixing} is to select an arbitrary copy in the first Gribov region, where the Faddeev-Popov operator is positive definite~\cite{Vandersickel:2012tz}. The minimal gauge-fixing strategy is the most popular choice in lattice calculations. The study in Ref.~\cite{Bornyakov:2011fn} evaluates three other different strategies: {\bf Simulated Annealing (SA), best copy}; {\bf SA, first copy} and {\bf Flipped Simulated Annealing (FSA), best copy}. SA is an algorithm used to improve the efficiency of the standard overrelaxation, and ``filpped" means applying $\mathbb{Z}(2)$ flip transformations on gauge configuration defined as
\begin{align}
f_\nu\left(U_{x, \mu}\right)=\left\{\begin{array}{cl}
-U_{x, \mu} & \text { if } \mu=\nu \text { and } x_\mu=a~, \\
U_{x, \mu} & \text { otherwise } ~.
\end{array}\right.
\label{eq:flip}
\end{align}
Accordingly, no ``filpped'' means that all copies are constrained in the first Gribov region. Besides, ``best copy" means choosing the maximum functional value among all copies, while ``first copy" chooses the first copy obtained by the FSA or SA algorithm.

More investigations on strategies to choose the Gribov copies in gauge-fixing can be found in~\cite{Parrinello:1990pm,Zwanziger:1990tn,Bogolubsky:2007bw,Bogolubsky:2009dc,Bornyakov:2011jm}. Within these approaches, the flip transformations $\mathbb{Z}(2)$ as defined in \Eq{flip}, which is first introduced in the context of Landau gauge-fixing in~\cite{Bogolubsky:2005wf}, is different from conventional methods of gauge-fixing. However, in the lattice calculation of the hadron structure functions, flip transformations are rarely used in gauge-fixing, so the $\mathbb{Z}(2)$ flip is not studied in this work. 

Considering the computational cost of the $3+1$ dimensional QCD simulation on the lattice, we study two strategies in this work: 
\begin{enumerate}
    \item {\bf First instance} (``First it"): choosing the first copy obtained in the first Gribov region;
    \item {\bf Smallest functional} (``Smallest f"): choosing the copy with the smallest functional value among all the obtained copies in the first Gribov region.
\end{enumerate}
Using these two strategies, we should obtain two different sets of configurations after fixing the CG. The measurement distortion introduced by Gribov copies is estimated via the comparison between these two sets.

\section{Numerical results}
\label{sec:numerical_results}

\begin{figure}
    \centering
    \includegraphics[width=0.9\linewidth]{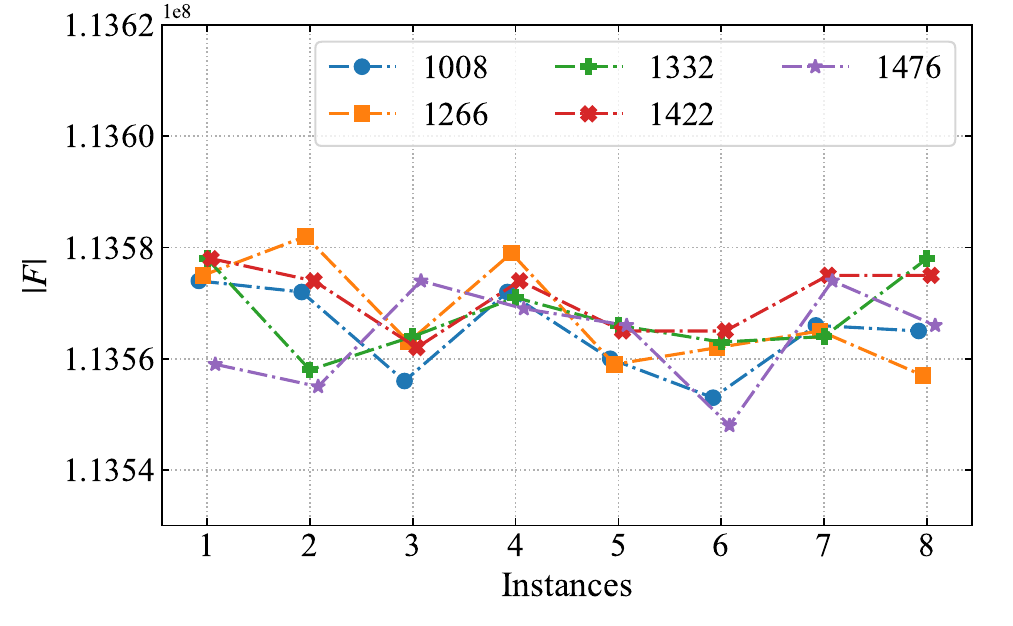}
    \caption{Distribution of functional values across five configurations, with the $x$-axis showing eight instances of gauge-fixing. 
    }
    \label{fig:func_val}
\end{figure}

To estimate the impact of Gribov copies, we perform a numerical lattice QCD calculation on a 2+1 flavor QCD gauge ensemble, generated by the HotQCD Collaboration~\cite{HotQCD:2014kol} with Highly Improved Staggered Quarks (HISQ)~\cite{Follana:2006rc}. The ensemble has a lattice spacing of $a=0.06$ fm and a volume of $L_s^3 \times L_t = 48^3 \times 64$. We used tadpole-improved clover Wilson valence fermions on a hypercubic (HYP) smeared~\cite{Hasenfratz:2001hp} gauge background. We set the clover coefficient $c_{\rm sw} = u_0^{-3/4}$, in which $u_0$ is the average plaquette after HYP smearing; the parameter is set as $c_{\rm sw} = 1.0336$ for both time and spatial directions. For the masses of the valence quarks, we used $\kappa \approx 0.12623$ for the light quarks $u$ and $d$, corresponding to a valence pion mass $m_\pi = 300$ MeV. We measured $100$ gauge configurations and effectively increased statistics $8$ times by using $1$ exact Dirac operator inversion (with precision $10^{-10}$) and $8$ sloppy inversions (with precision $10^{-4}$) on each of the configurations using All-Mode Averaging (AMA)~\cite{Shintani:2014vja}. The details of the measurements are collected in \tb{setup}. Using this setup, the statistical uncertainties associated with the matrix elements are maintained at subpercent level, which is sufficient for the examination of the systematic uncertainties from Gribov copies. Using the mother-daughter methodology referenced in Sec.~\ref{sec:method}, we employ the Steepest Descent algorithm to obtain $8$ instances of functional minima on each configuration. In the case of minimal functional $F$, the determinant of the Faddeev-Popov operator is non-negative, so all our instances are in the first Gribov region~\cite{Giusti:2001xf}. A recent paper~\cite{Zhang:2024omt} shows that gauge-dependent measurements are very sensitive to gauge-fixing precision. Only when a high precision is achieved can the continuum limit be observed between results from different lattice ensembles, as in our previous work in \refcite{Gao:2023lny}. Therefore, in this work, we choose a high precision of $\delta F / F \leq 10^{-8}$ for gauge-fixing on each configuration. Different copies are distinguished by different functional values that exceed the precision of the gauge-fixing. Two sets of Gribov copies are selected on $100$ configurations according to the strategies mentioned in Sec.~\ref{sec:method}, named ``First it" and ``Smallest f", respectively.

\begin{table}[h]
\renewcommand{\arraystretch}{1.2} 
    \begin{center}
    \begin{tabular}{|c|c|c|c|c|c|c|}
        \hline
        \hline
        $a$ & $L_s^3 \times L_t$ & $m_q a$ & $m_\pi L_t$ & \#Cfgs & \#It & (\#ex, \#sl) \\
        \hline
        $0.06$~fm & $48^3 \times 64$ & -0.0388 & 5.85 & 100 & 8 & (1, 8) \\
        \hline
        \hline
    \end{tabular}
    \end{center}
    \caption{A comprehensive overview of the lattice configuration and measurement parameters is provided. The symbol \#Cfgs denotes the number of configurations, while \#It represents the frequency of gauge-fixing instances per configuration. Furthermore, \#ex and \#sl indicate the counts of precise inversions and approximate inversions, respectively. We have specified the bare Wilson fermion quark mass $m_q a$ corresponding to a $300$~MeV pion mass $m_\pi$, the temporal extent $L_t$ of the lattice in $m_\pi$ units.}
    \label{tb:setup}
\end{table}

As illustrated in Fig.~\ref{fig:func_val}, five configurations (configuration numbers: 1008, 1266, 1332, 1422, and 1476) are randomly selected. The absolute values $|F|$ of the functional, as defined in Eq.~(\ref{eq:functional}), are plotted following each gauge-fixing instance. Given that the precision of gauge-fixing is set to $\delta F / F \leq 10^{-8}$, when the difference in functional value between two instances exceeds $F \times 10^{-8}$, they are identified as distinct Gribov copies. As shown in Fig.~\ref{fig:func_val}, most instances belong to distinct copies. Therefore, our strategy of selecting the different Gribov copies works effectively in these configurations.

\subsection{Quark propagator}

Most of the lattice QCD calculations start from the measurements of correlations. In order to assess the impact of Gribov copies in general, we measured light-quark propagators at different separations $z$ with a point source and a point sink in the coordinate space. Since the CG is independently fixed on each time slice, light-quark propagators are chosen along a spatial direction to have a non-vanishing signal. The light-quark propagator is defined as 
\begin{align}
    C_u (z) = \langle \mathrm{Tr} [ u(z) \bar{u}(0) ] \rangle ~,
\end{align}
where $\mathrm{Tr}$ denotes the trace over both color and spin indices. The decaying behavior of $C_u (z)$ can be presented more clearly with the definition of an effective mass $m_\mathrm{eff}$ of the dressed quark at each $z$ as~\cite{Gattringer:2010zz} through
\begin{align}
    \frac{C_u (z)}{C_u (z+1)} = \frac{ \cosh(m_\mathrm{eff} \cdot (z - L_s / 2) ) }{ \cosh(m_\mathrm{eff} \cdot (z + 1 - L_s / 2) ) } ~,
\end{align}
where $L_s$ is the lattice size in spatial directions. 
\begin{figure}[htbp]
    \centering
    \includegraphics[width=.4\textwidth]{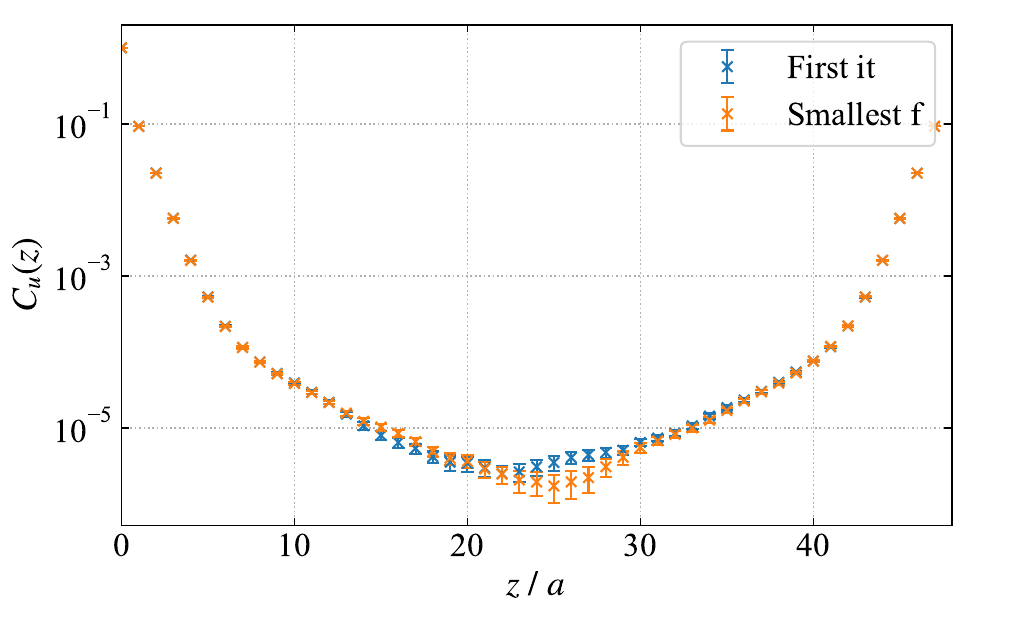}
    \includegraphics[width=.4\textwidth]{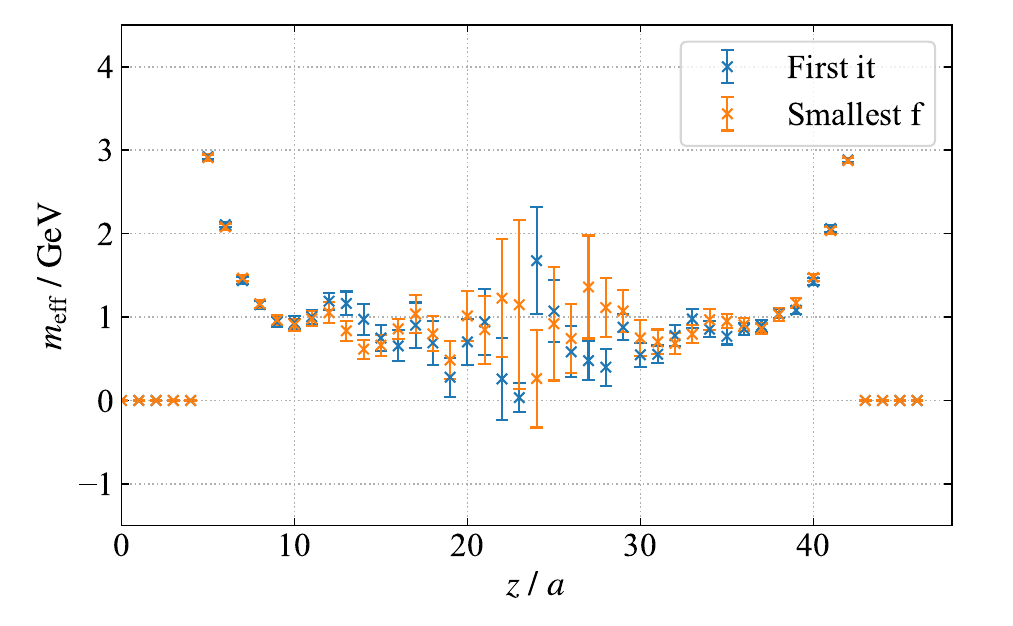}
    \caption{Two-point correlation functions(up) and effective mass of the quark spatial propagator(down) using two different strategies are plotted. The uncertainties are estimated using bootstrap resampling on $100$ configurations. \label{fig:qpz_meff_N100}}
\end{figure}
Both the two-point correlation functions and the effective masses of the two gauge-fixing strategies are shown in Fig.~\ref{fig:qpz_meff_N100}, with the uncertainties estimated using bootstrap resampling. Good consistency between two sets can be found in the plots, which indicates that the measurement distortions from Gribov copies are smaller than the statistical noise.

\begin{figure}
    \centering
    \includegraphics[width=.42\textwidth]{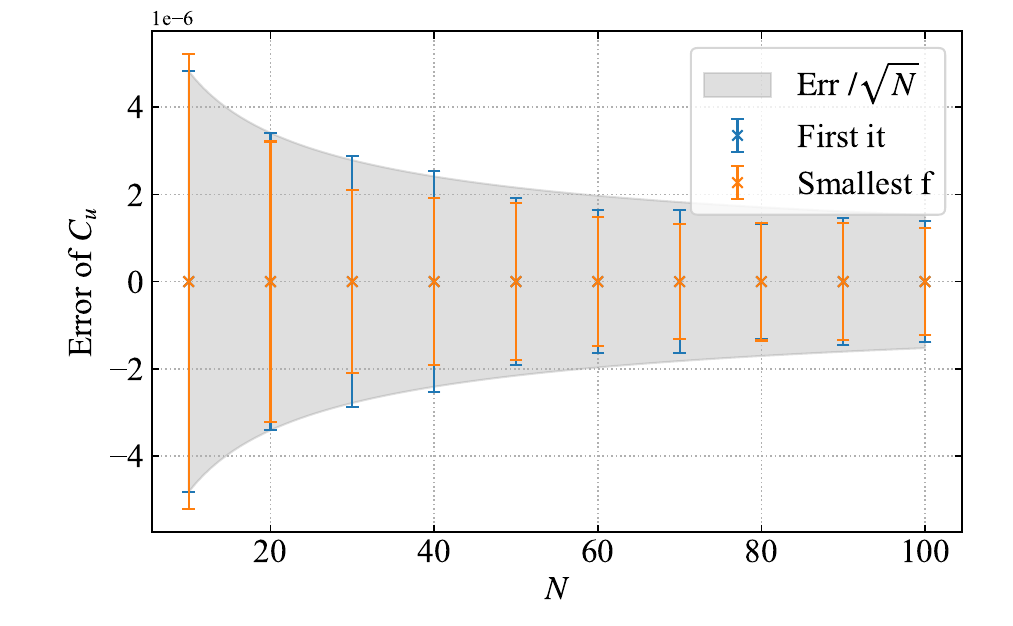}
    \includegraphics[width=.42\textwidth]{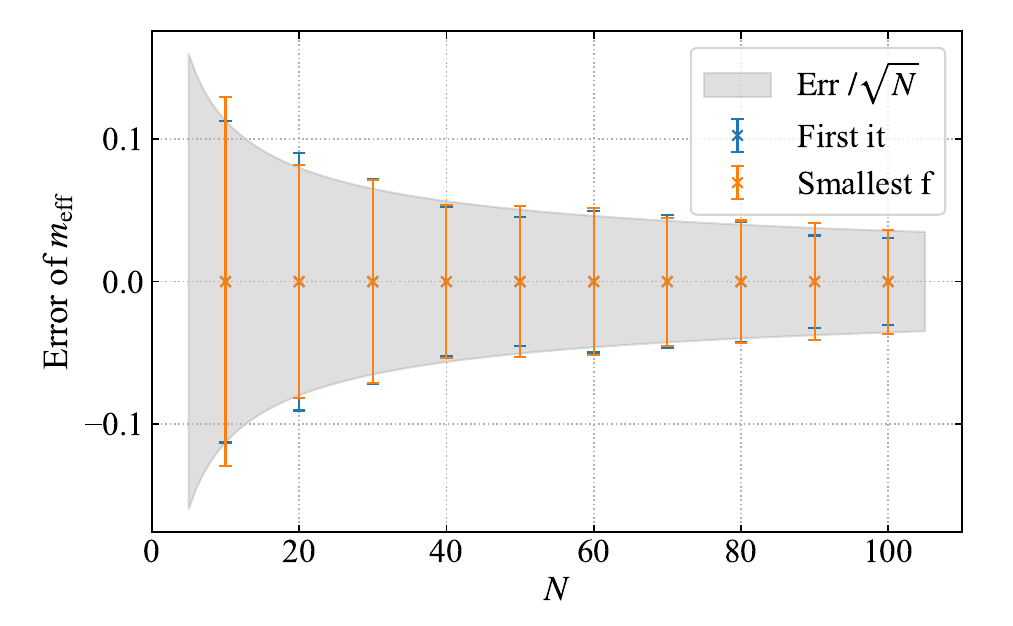}
    \caption{The errors of the quark propagator and the effective mass at $z=10a$ as a function of number of configurations $N$. The grey band shows the expected error that decreases as $1/\sqrt{N}$.\label{fig:qpz_err}}
\end{figure}

Considering that an imprecise gauge-fixing or a skewed gauge-fixing strategy could result in a complex and irregular distribution of the target quantity across different configurations and Gribov copies, the error estimation with a finite number of configurations requires careful evaluation. To verify that our number of configurations is sufficient to give a correct error estimation, the errors of $C_u (z)$ and $m_\mathrm{eff}(z=10a)$ are plotted as functions of the number of configurations $N$ in Fig.~\ref{fig:qpz_err}. It is well known that statistical uncertainties decrease according to $1/\sqrt{N}$ as the number of configurations $N$ increases. Thus, observing a consistent pattern suggests that our bootstrap samples are sufficient to describe the distribution of quark spatial propagator across different configurations and Gribov copies, and increasing the number of measurements can reach higher precision. Moreover, this serves as supporting evidence that our gauge-fixing precision is sufficiently high. The gray bands in the figures show the expected error as $C_u (N=10) / \sqrt{N}$ and $m_\mathrm{eff}(N=10) / \sqrt{N}$ respectively. The data from both ``First it" and ``Smallest f" align well with the expected error; this indicates that the errors generally behave as statistical noise with negligible systematic uncertainties from Gribov copies.

\subsection{Quasi-distribution matrix elements}

To further investigate the systematic uncertainties of Gribov copies, specifically in the lattice calculation of parton physics, we calculate the spatial correlators for pion quasi-distributions using the two group of Gribov copies mentioned above. Due to the 3D rotational invariance under the CG, the correlators can be used to define both PDFs~\cite{Gao:2023lny} and TMDs~\cite{Zhao:2023ptv,Bollweg:2024zet}.

The bare quasi-distribution matrix element under the CG is~\cite{Gao:2023lny}
\begin{align}
    \tilde{h}_{\gamma^t}(z, P^z, \mu) = \frac{1}{2 P^t} \langle P|\bar{\psi}(z) \gamma^t \psi(0)|_{\vec{\nabla} \cdot \vec{A}=0}|P \rangle ~,
 \label{eq:quasi_pdf}
\end{align}
which can be extracted from the ratio of three-point correlations and two-point correlations (see \app{baremx}). Without loss of generality, we focus on the $P^z=0$ matrix elements of pion that are the least computationally expensive. Three time separations $t_{\rm sep} = \{8, 10, 12\}a$ are measured for the three-point correlations $C_{\rm 3pt}(t_{\rm sep})$. To eliminate excited-state contamination, we fit the two-point correlations $C_{\rm 2pt}(t)$ with two energy states in $t \in [3, 15]$. Then the fit posteriors are passed onto the two-state fit of the ratio $R(t_{\rm sep}, \tau) = C_{\rm 3pt}(t_{\rm sep}, \tau) / C_{\rm 2pt}(t_{\rm sep})$ as priors, in which $\tau$ is the insertion time of the operator. All the fits are least square fits based on Bayesian analysis. Bootstrap resampling is adopted to establish correlations among all data points, and the correlations are maintained consistently throughout the analysis. 

\begin{figure}[htbp]
\centering
\includegraphics[width=.4\textwidth]{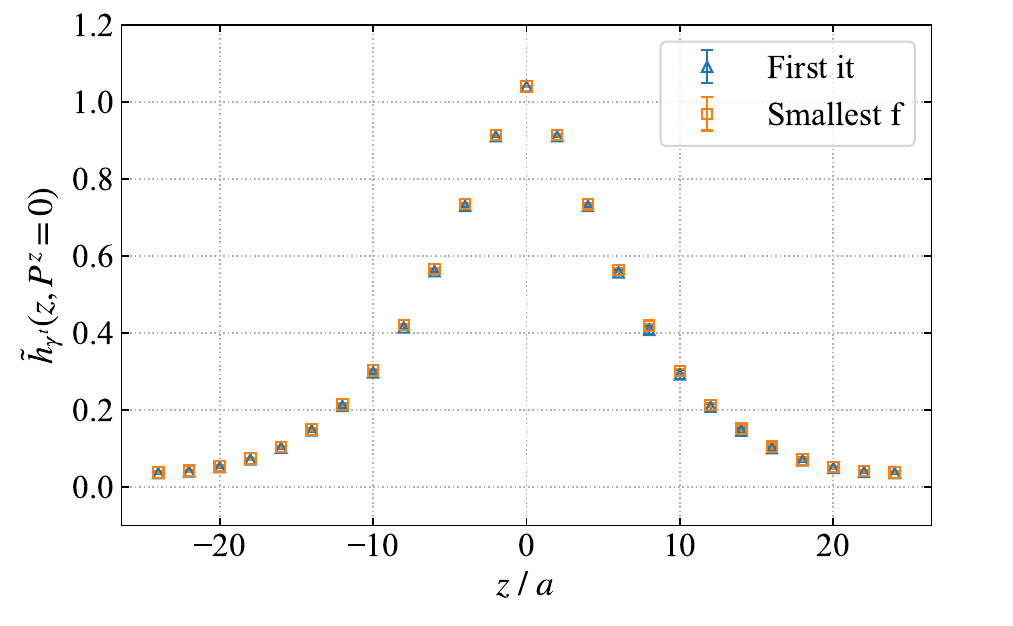}
\includegraphics[width=.4\textwidth]{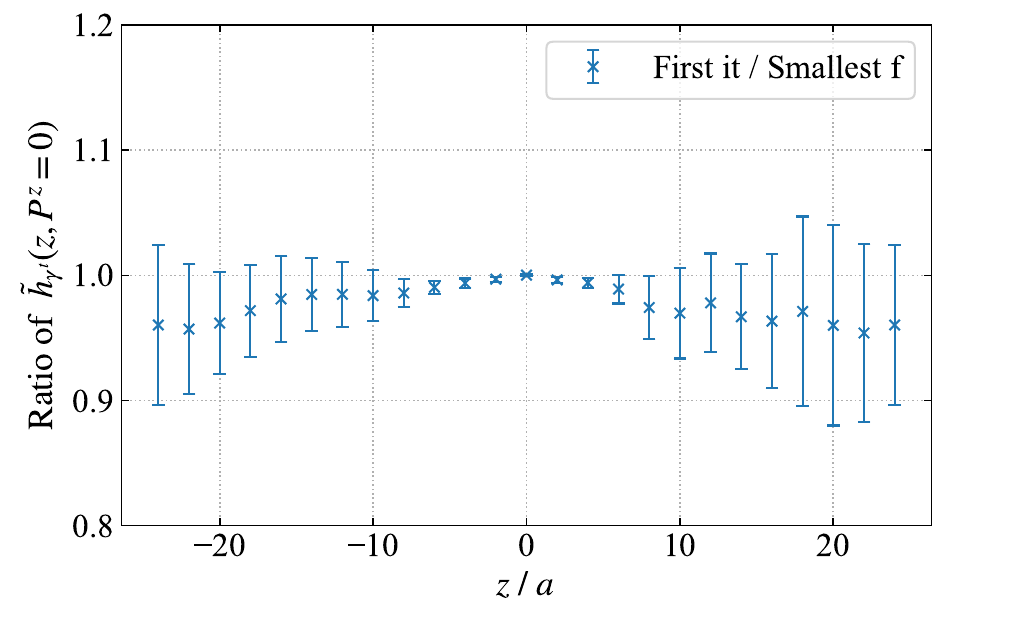}
\caption{Upper panel: The $z$-dependence of the pion quasi-distribution matrix element using each strategy. Lower panel: The $z$-dependence of the ratio of matrix elements from the two strategies.\label{fig:z_dep_ratio}}
\end{figure}

\begin{figure*}[htbp]
\centering
\includegraphics[width=.4\textwidth]{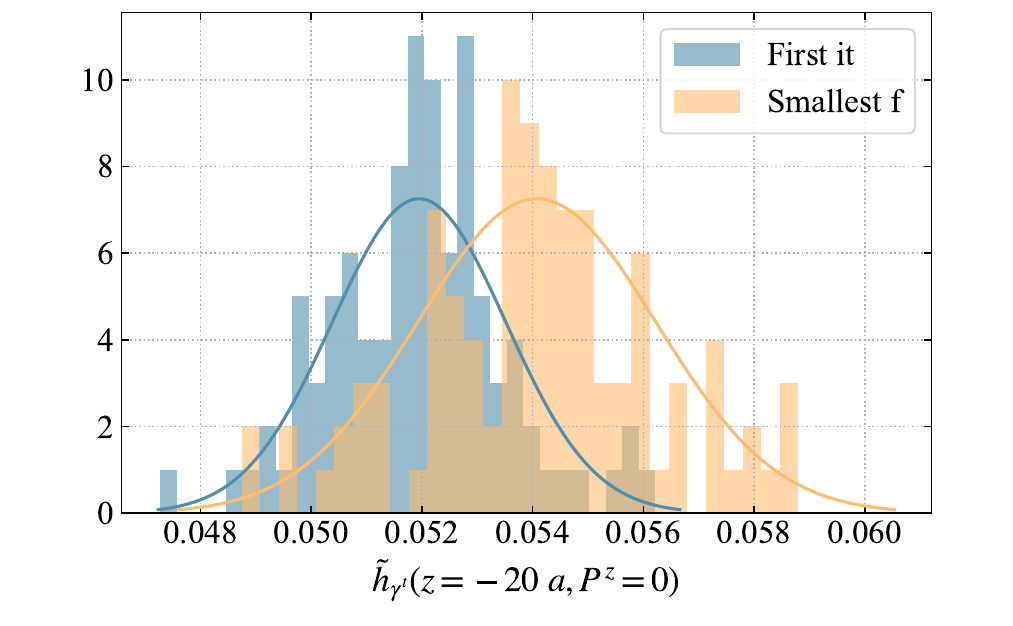}
\includegraphics[width=.4\textwidth]{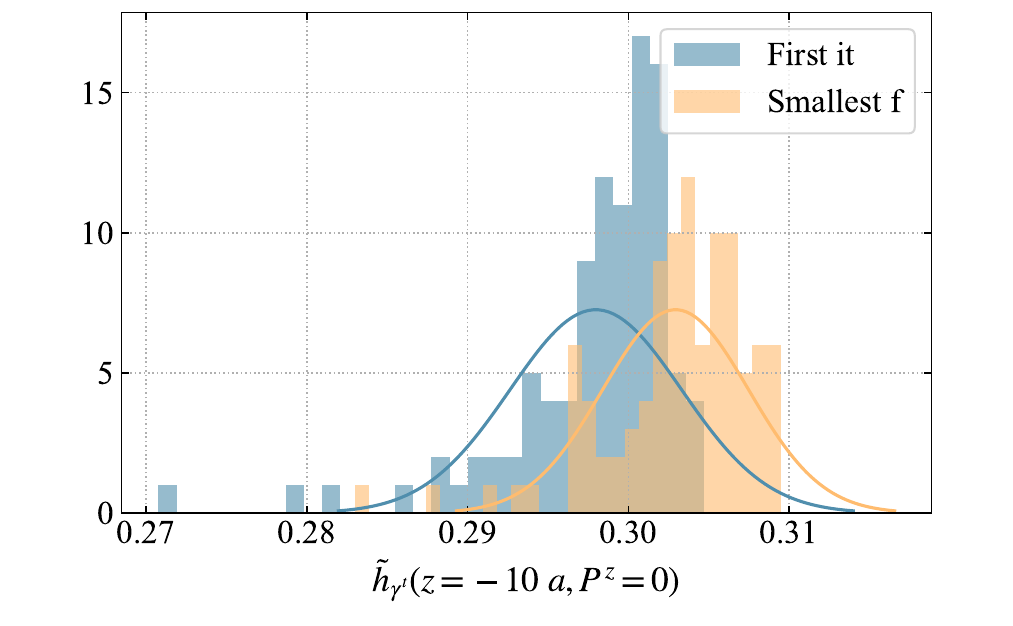}
\includegraphics[width=.4\textwidth]{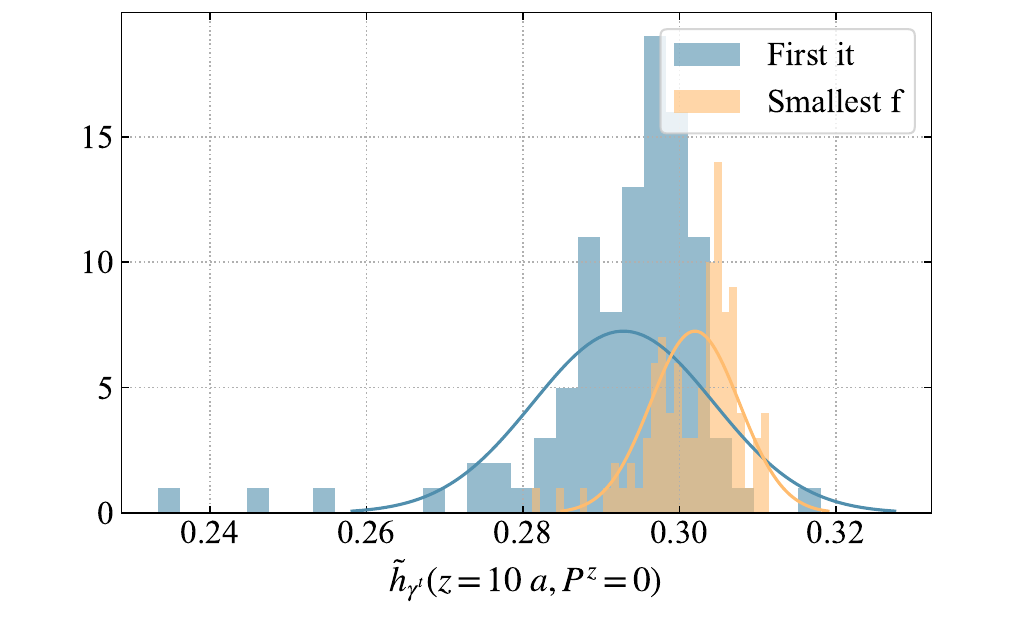}
\includegraphics[width=.4\textwidth]{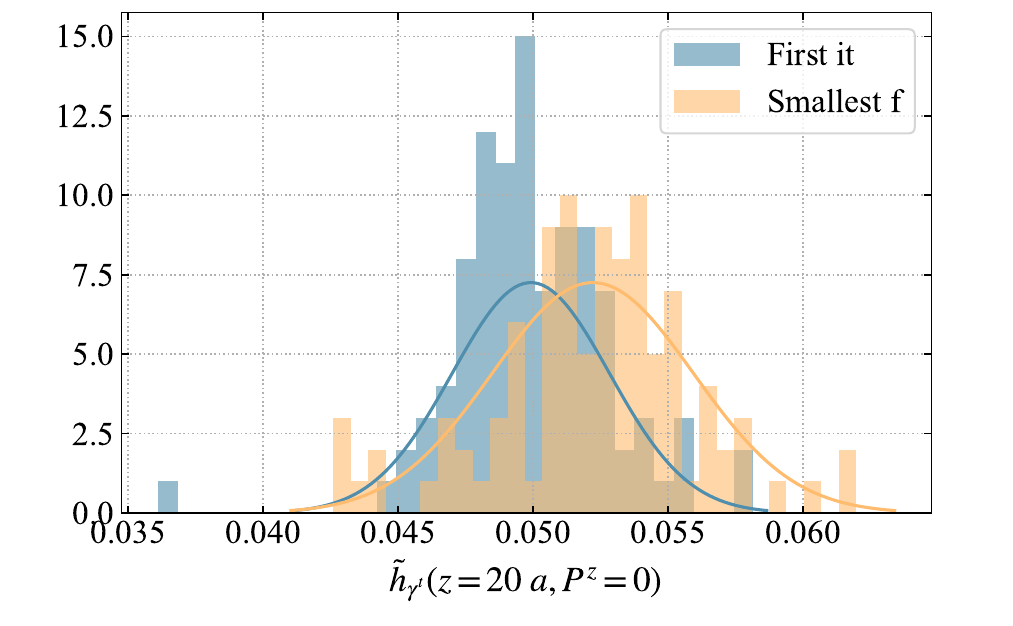}
\caption{Histograms of the matrix element $\tilde h_{\gamma^t}(z, P^z=0)$ from the bootstrap samples of the 100 configurations, measured on the two different sets of Gribov copies.\label{fig:histogram}}
\end{figure*}

To compare calculations with the two different strategies, the bare matrix elements and their ratio are plotted in Fig.~\ref{fig:z_dep_ratio}, which show good consistency within $1 \sigma$.
In order to illustrate the statistical comparison between the two sets in detail, we select four random separations $z=\{-20, -10, 10, 20\} a$ and plot histograms from bootstrap samples of the matrix elements in Fig.~\ref{fig:histogram}. The matrix elements show no statistical difference between two strategies for choosing Gribov copies in the gauge-fixing process, suggesting that the measurement distortion of the Gribov copies is smaller than the statistical uncertainties.

\begin{figure*}[htbp]
\centering
\includegraphics[width=.4\textwidth]{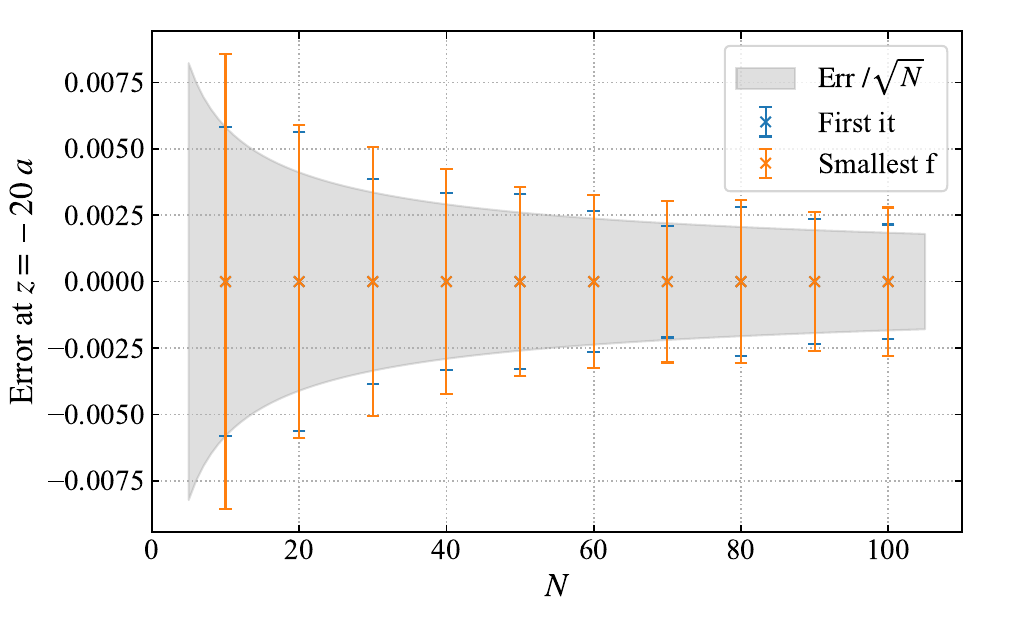}
\includegraphics[width=.4\textwidth]{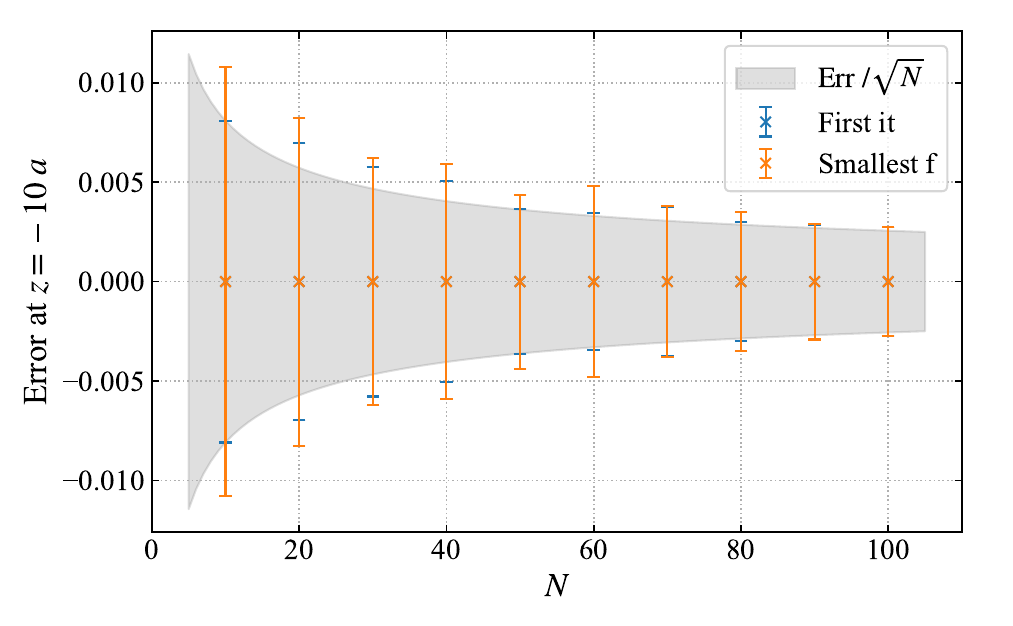}
\includegraphics[width=.4\textwidth]{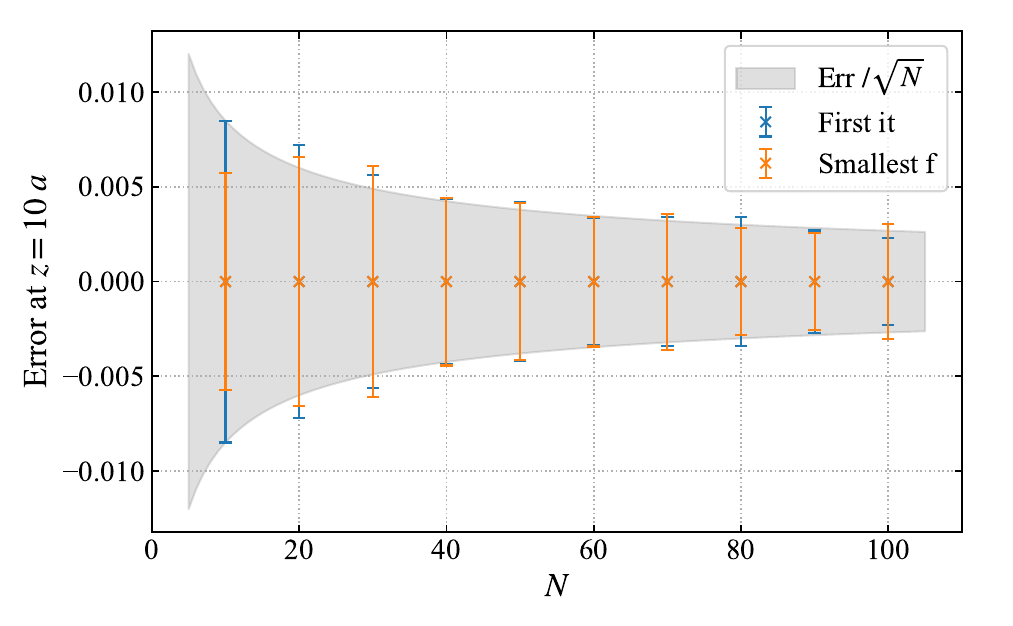}
\includegraphics[width=.4\textwidth]{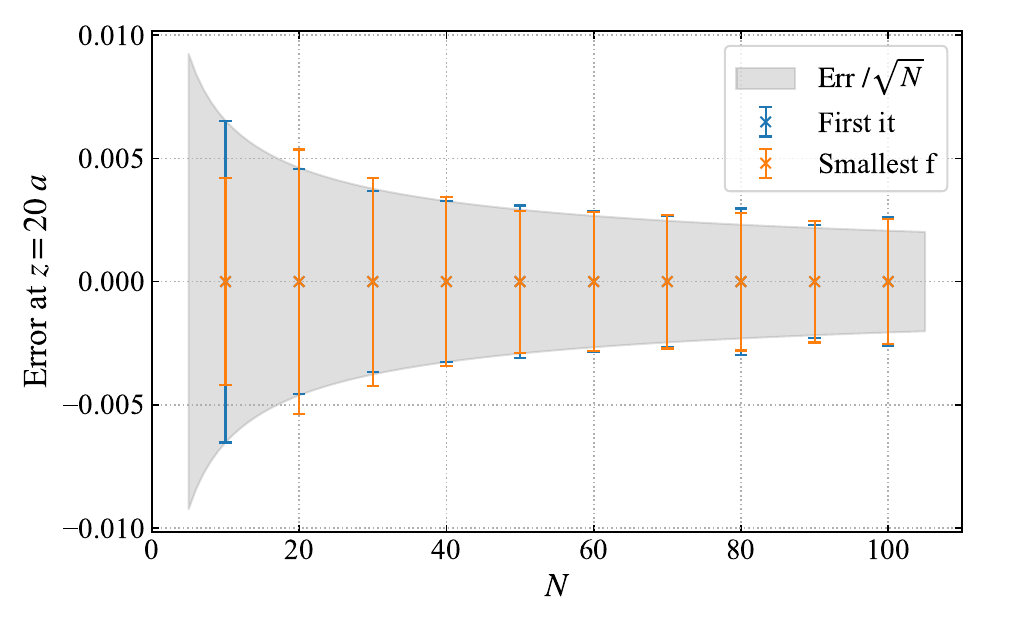}
\caption{The error of ratio at different separations $z$, $t_{\rm sep} = 8a$ and $\tau = 4a$. Varying $N$ corresponds to using different numbers of configurations, each resampled via bootstrap to get 100 samples. \label{fig:ratio_err}}
\end{figure*}

Analogous to Fig.~\ref{fig:qpz_err}, the assessment of our error estimation can be conducted by examining the variation in uncertainties relative to the number of configurations. We plot the error of the ratio $R(t_{\rm sep} = 8a, \tau=4a)$ for various separations $z$ in Fig.~\ref{fig:ratio_err}. The grey band shows the expected error as $R(N = 10) / \sqrt{N}$, which is in good agreement with both sets of Gribov copies. It shows that the error estimation for both the ``First iteration" and the ``Smallest functional" as defined in Sec.~\ref{sec:method}, are reliable.

\section{Conclusion}
\label{sec:conclusions}

In conclusion, this work assessed the effect of Gribov copies in the measurements of quark correlators in the CG via the comparison of two different strategies (``First it" and ``Smallest f") in gauge-fixing. Our results on the quark propagator and the quasi-distribution matrix elements show that the systematic uncertainties from Gribov copies, i.e. measurement distortions, are negligible compared with the statistical uncertainties in the CG matrix elements. It is consistent with the argument in~\refcite{Gao:2023lny} and supports the effectiveness of the CG method in the calculation of parton physics. 

In the future, more instances of gauge-fixing, i.e. more Gribov copies, can be included to examine the full extent of measurement distortion. The recent study in \refcite{Kalusche:2024osk} deployed 100 copies to analyze the influence of Gribov copies on the quark propagator, though limited to two colors. Furthermore, other strategies can be explored to select diverse Gribov copies to strengthen our conclusion. Moreover, the effect of Gribov copies on CG gluon correlators, which has been studied for gluon propagators in SU(2) gauge theory~\cite{Maas:2008ri,Bornyakov:2011fn}, will be investigated in SU(3) QCD for their application to the gluon parton distributions.

\begin{acknowledgments}
The measurement of the correlators was carried out with the \texttt{Qlua} software suite~\cite{qlua}, which utilized the multigrid solver in \texttt{QUDA}~\cite{Clark:2009wm,Babich:2011np}. Part of the calculations were performed using the \texttt{Grid} \cite{Boyle:2016lbp,Yamaguchi:2022feu} and \texttt{GPT} \cite{GPT} software packages.

We thank George Flemming, Xiangdong Ji, Peter Petreczky, Swagato Mukherjee and Dennis Bollweg for valuable communications. This material is based upon work supported by the U.S. Department of Energy, Office of Science, Office of Nuclear Physics through Contract No.~DE-AC02-06CH11357, and within the frameworks of Scientific Discovery through Advanced Computing (SciDAC) award \textit{Fundamental Nuclear Physics at the Exascale and Beyond} and the Quark-Gluon Tomography (QGT) Topical Collaboration, under contract no.~DE-SC0023646. 
YZ is also partially supported by the 2023 Physical Sciences and Engineering (PSE) Early Investigator Named Award program at Argonne National Laboratory.

This research used awards of computer time provided by the INCITE program at Argonne Leadership Computing Facility, a DOE Office of Science User Facility operated under Contract DE-AC02-06CH11357, the ALCC program at the Oak Ridge Leadership Computing Facility, which is a DOE Office of Science User Facility supported under Contract DE-AC05-00OR22725, the National Energy Research
Scientific Computing Center, a DOE Office of Science User Facility supported by the Office of Science of the U.S. Department of Energy under Contract DE-AC02-05CH11231 using NERSC award NP-ERCAP0028137. Computations for this work were carried out in part on facilities of the USQCD Collaboration, which is funded by the Office of Science of the U.S. Department of Energy.
\end{acknowledgments}

\appendix

\section{Ground state fit}\label{app:baremx}

As mentioned in the main text, we have performed a fully correlated Bayesian analysis of the two-point and three-point correlation functions to extract the bare matrix element $\tilde{h}^0_{\gamma^t}(z, P^z = 0)$. The correlation is taken into account by performing a Bayesian least squares fit on each sample of bootstrap. The parameter settings for the ground state fit are collected in \tb{gsfit}.
\begin{table}[h]
\renewcommand{\arraystretch}{1.2}
    \centering
    \begin{tabular}{|c|c|c|c|}
        \hline\hline
        Fit & \#state & $t_{\rm sep}$ range & $\tau$ range \\
        \hline
        $C_{\rm 2pt} (t_{\rm sep})$ & 2 & $t_{\rm sep} \in [3, 15]$ & N/A \\
        \hline
        $R(t_{\rm sep}, \tau)$ & 2 & $t_{\rm sep} \in \{ 8, 10, 12 \}$ & $\tau \in [3, t_{\rm sep} - 3]$ \\
        \hline\hline
    \end{tabular}
    \caption{Collection of ground state fit settings. \#state = 2 means there are 1 ground state and 1 excited state in the fit functions.}
    \label{tb:gsfit}
\end{table}

\begin{figure}[h]
    \centering
    \includegraphics[width=0.4\textwidth]{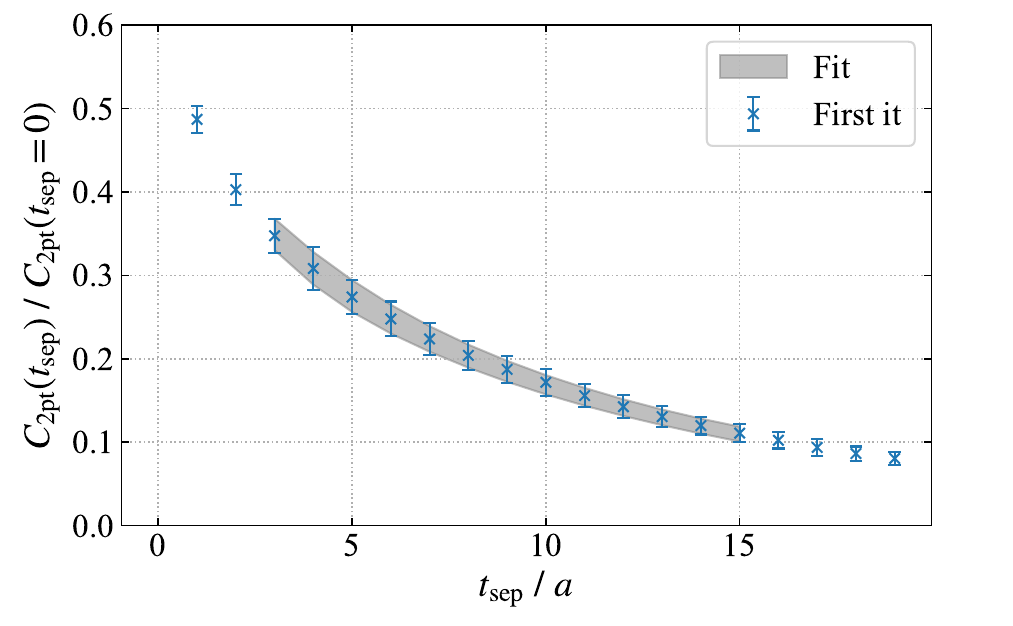}
    \includegraphics[width=0.4\textwidth]{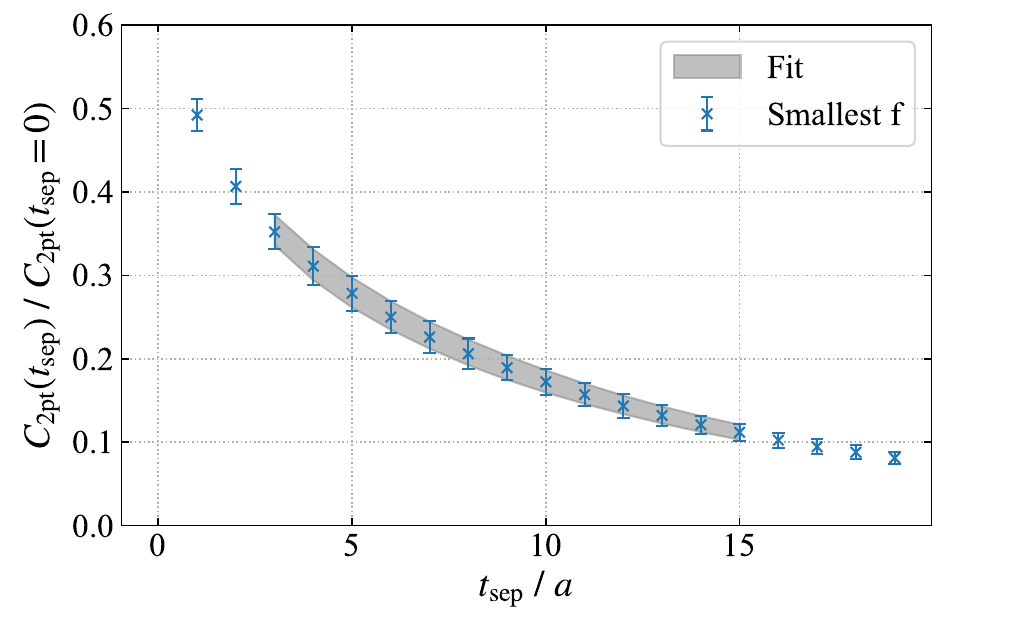}
    \caption{Two-point function of the pion non-local quark correlator; the uncertainties of the data points and the fit results are magnified by a factor of ten to facilitate a clearer visualization.}
    \label{fig:gsfit_2pt}
\end{figure}

Two-point correlation functions are fit using the function as
\begin{widetext}
\begin{align}
    C_{\rm 2pt} (t_{\rm sep}) = z_0^2 \left( e^{-E_0 t_{\rm sep}} + e^{-E_0 (L_t - t_{\rm sep})} \right) + z_1^2 \left( e^{-E_1 t_{\rm sep}} + e^{-E_1 (L_t - t_{\rm sep})} \right) ~,
\end{align}
\end{widetext}
in which $L_t$ is the length of the lattice in the time direction. The fit results and data points are plotted in \fig{gsfit_2pt}, where the uncertainties of the data points and the fit results are magnified by a factor of ten to facilitate a clearer visualization. The fit results demonstrate a high degree of consistency with the two-point functions.

\begin{figure*}[h]
\centering
\includegraphics[width=.4\textwidth]{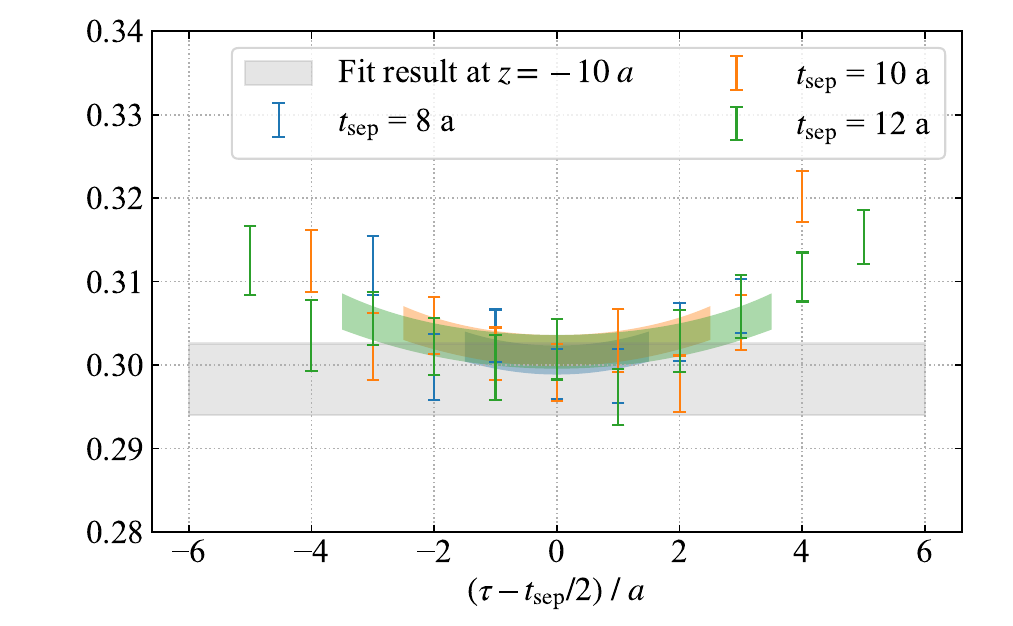}
\includegraphics[width=.4\textwidth]{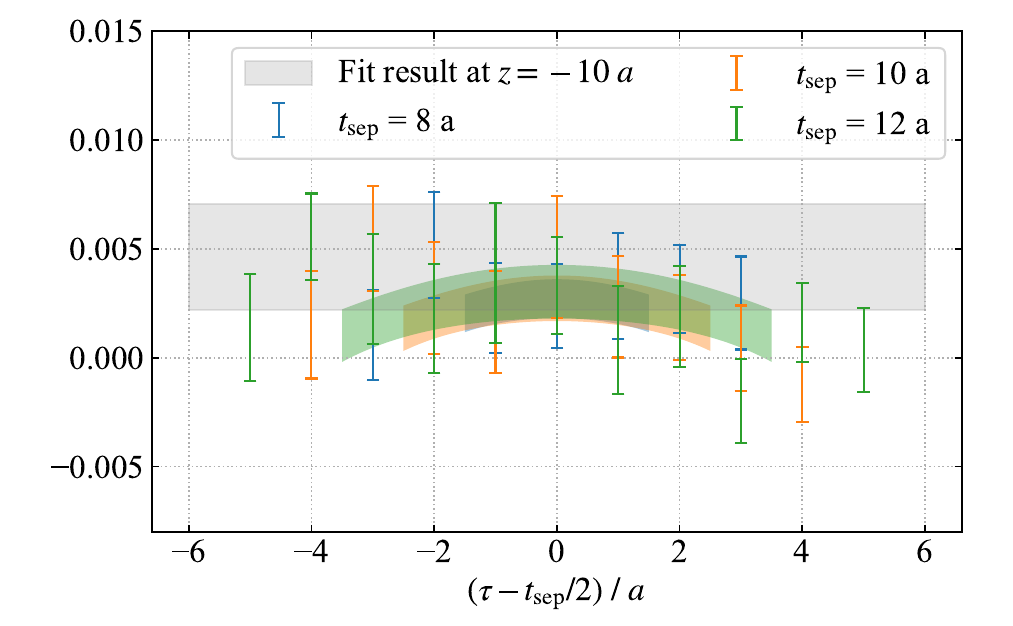}
\includegraphics[width=.4\textwidth]{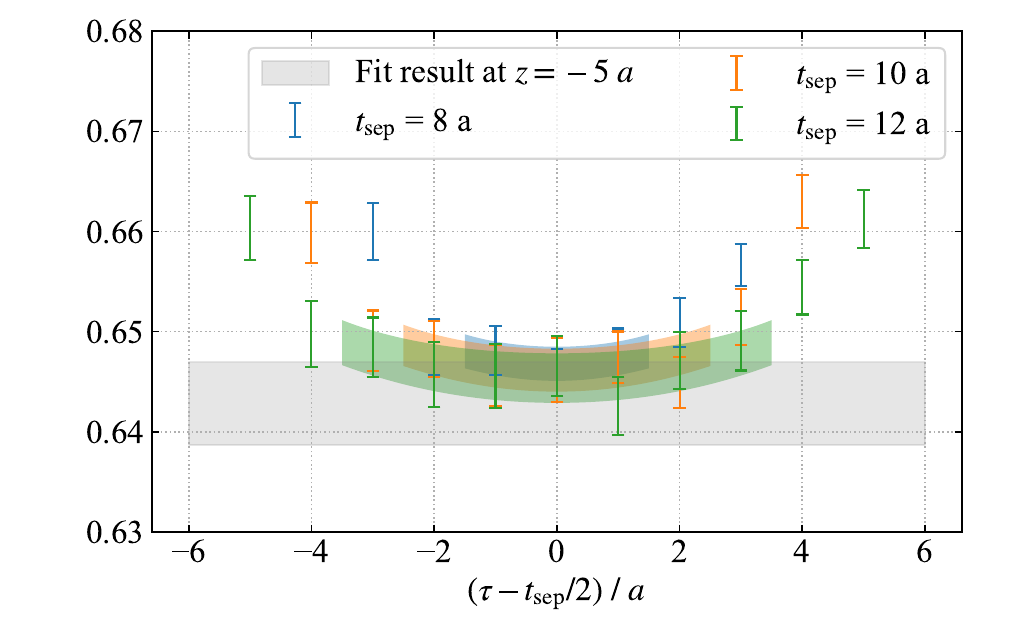}
\includegraphics[width=.4\textwidth]{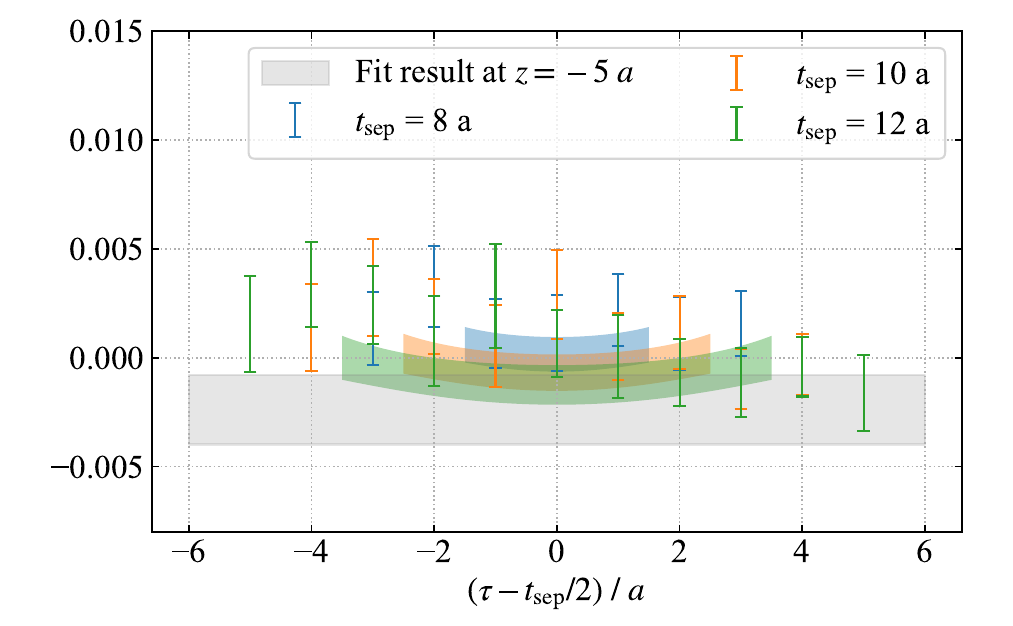}
\includegraphics[width=.4\textwidth]{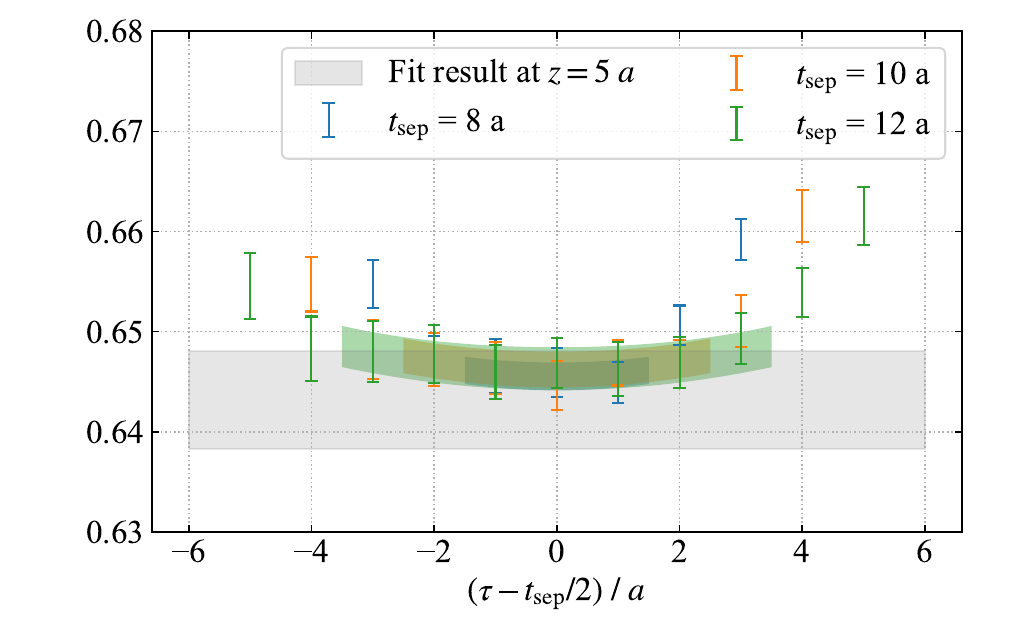}
\includegraphics[width=.4\textwidth]{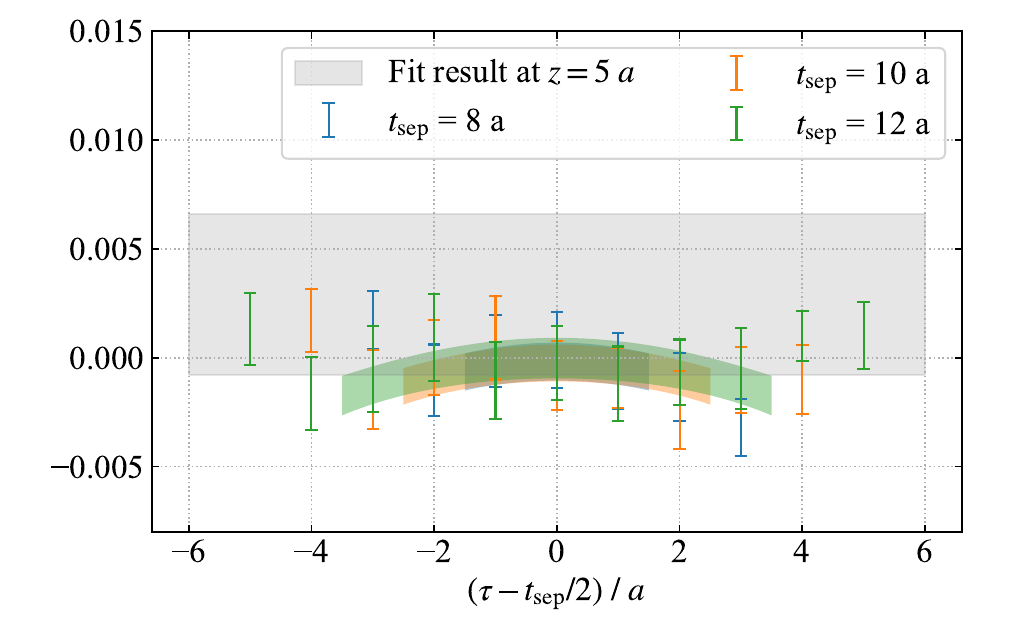}
\includegraphics[width=.4\textwidth]{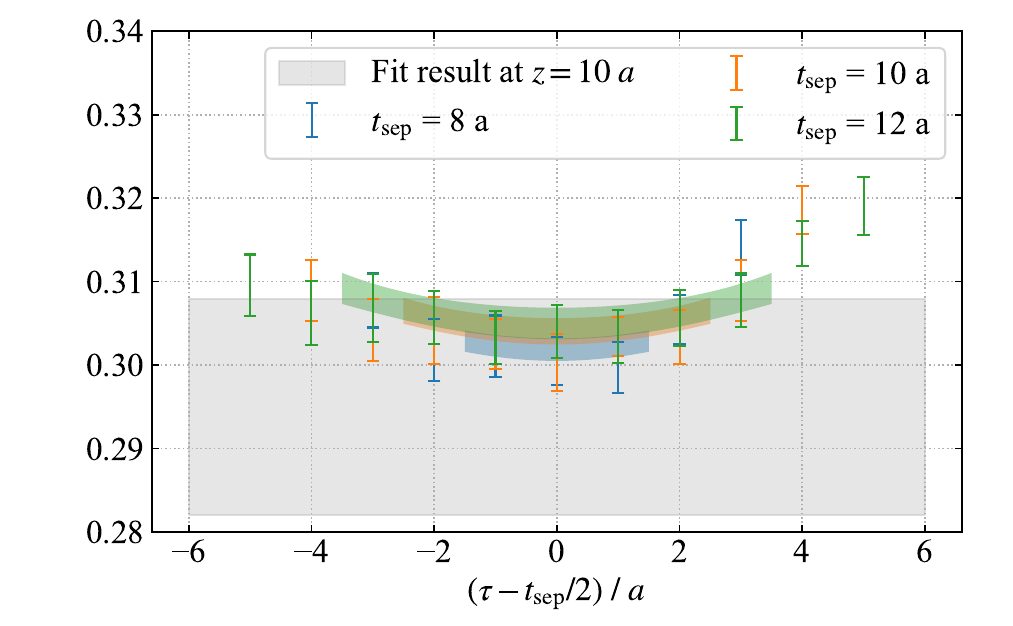}
\includegraphics[width=.4\textwidth]{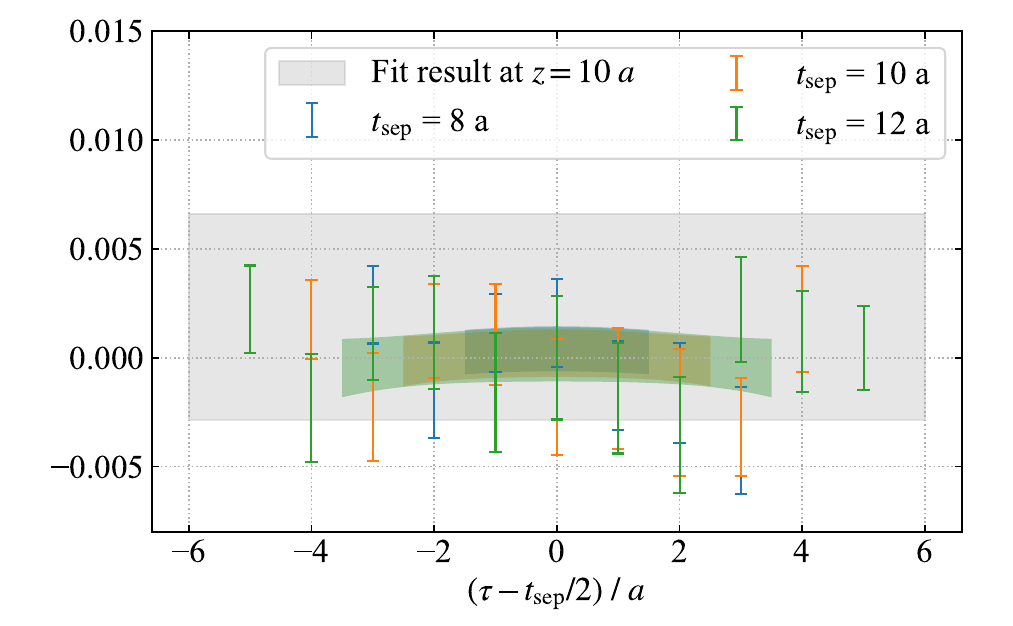}
\caption{Ratio fit of the pion non-local quark correlator using First it. Left column is the real part of the matrix elements and right column is imagnary part of the matrix elements. \label{fig:gsfit_ratio_first}}
\end{figure*}

\begin{figure*}[h]
\centering
\includegraphics[width=.4\textwidth]{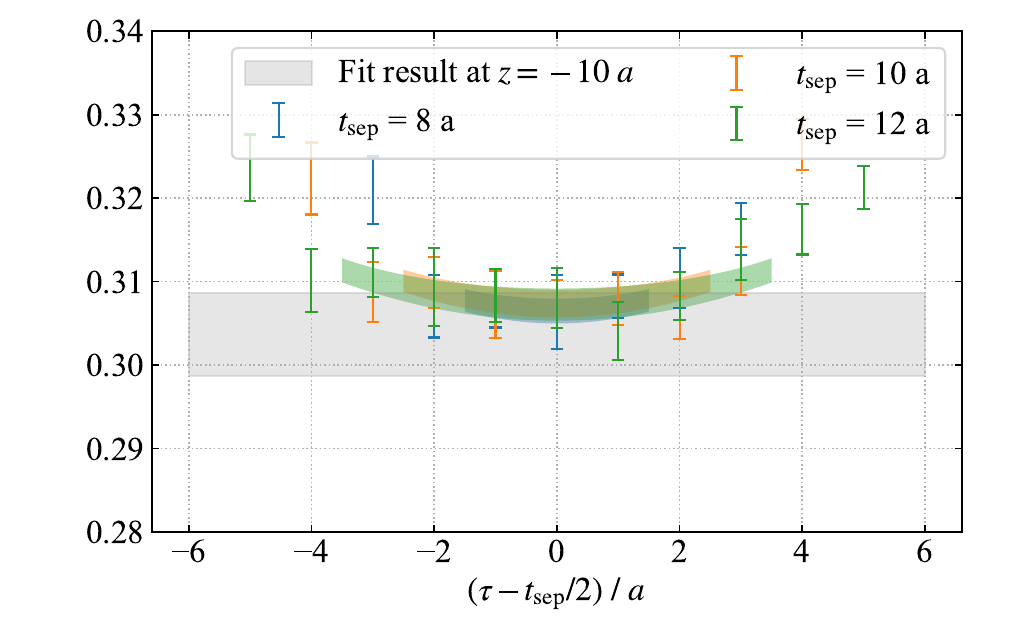}
\includegraphics[width=.4\textwidth]{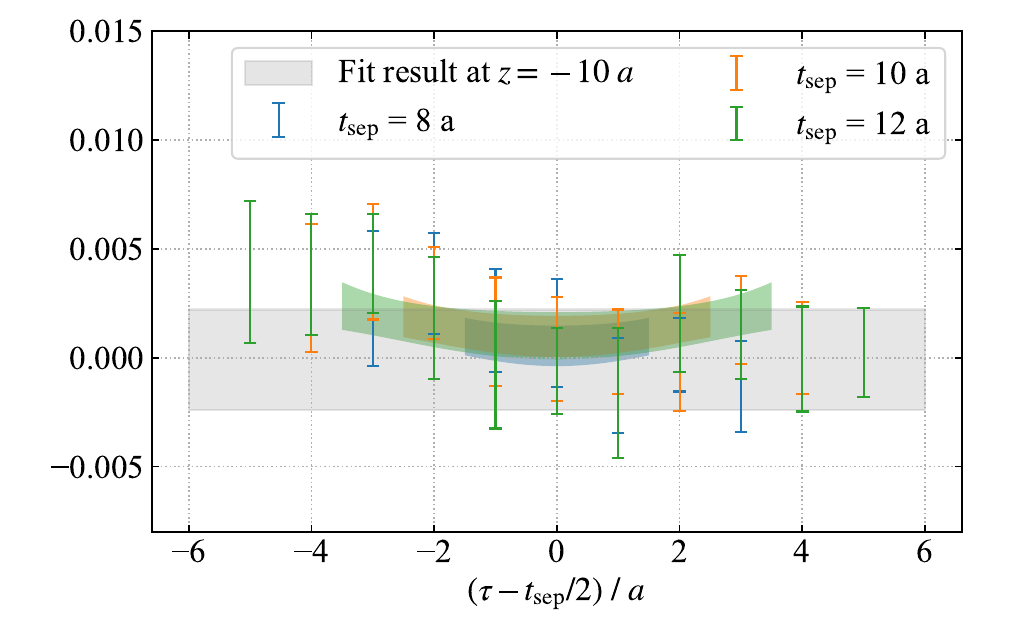}
\includegraphics[width=.4\textwidth]{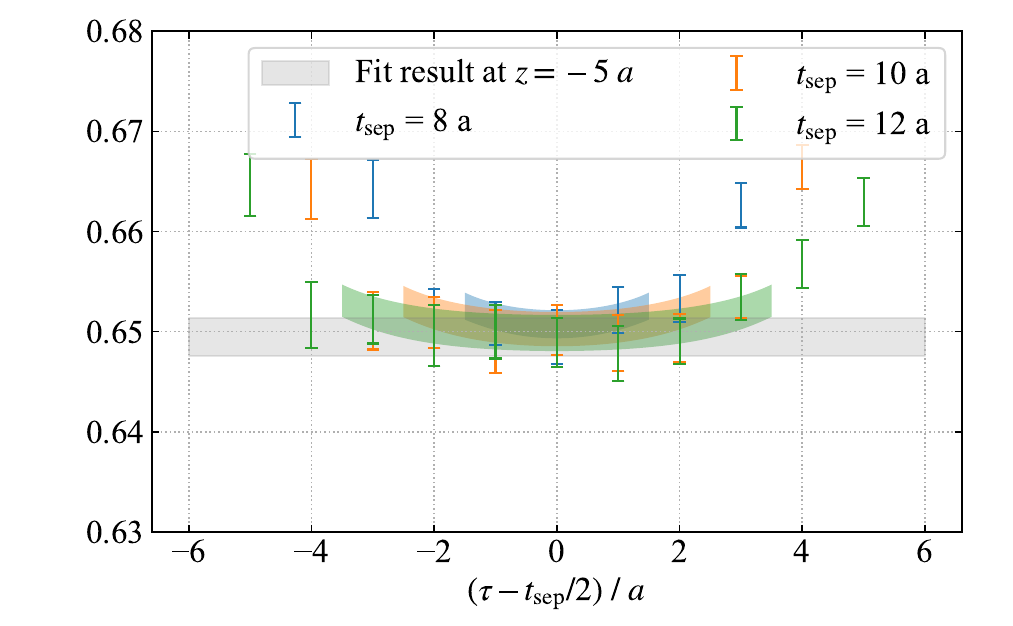}
\includegraphics[width=.4\textwidth]{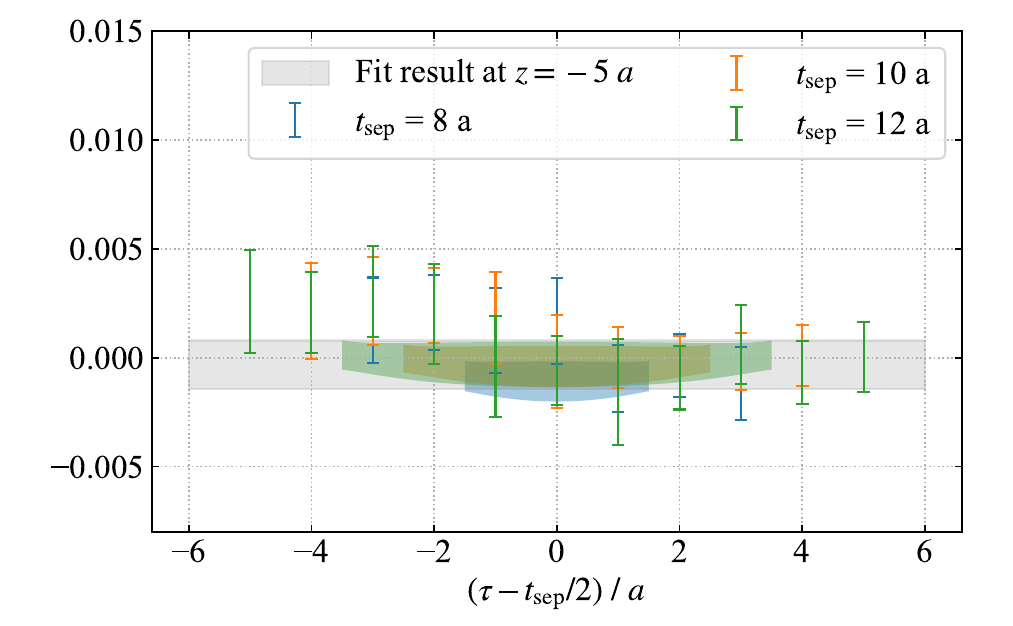}
\includegraphics[width=.4\textwidth]{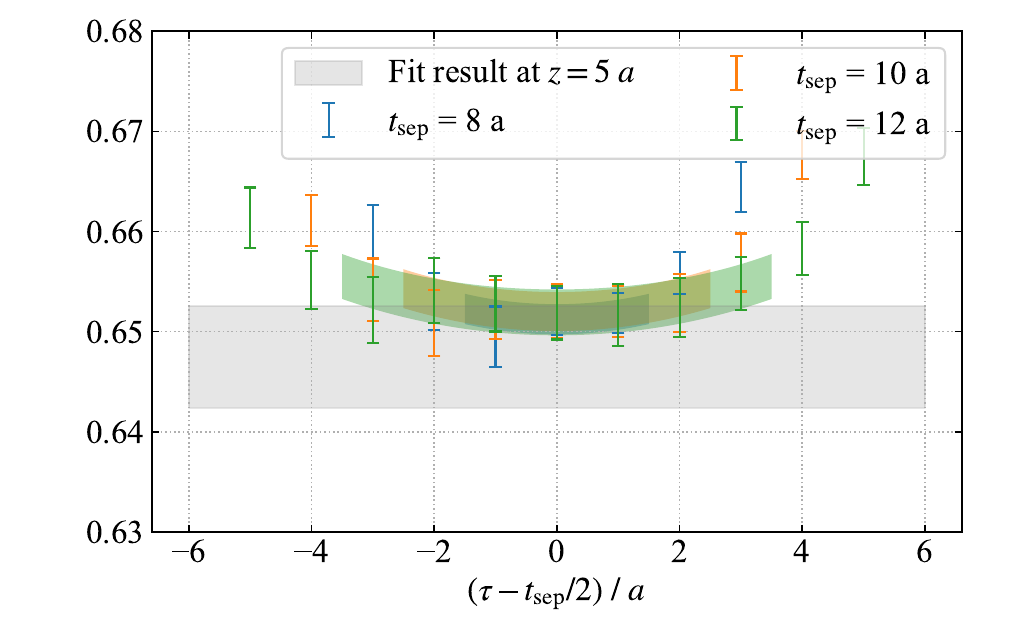}
\includegraphics[width=.4\textwidth]{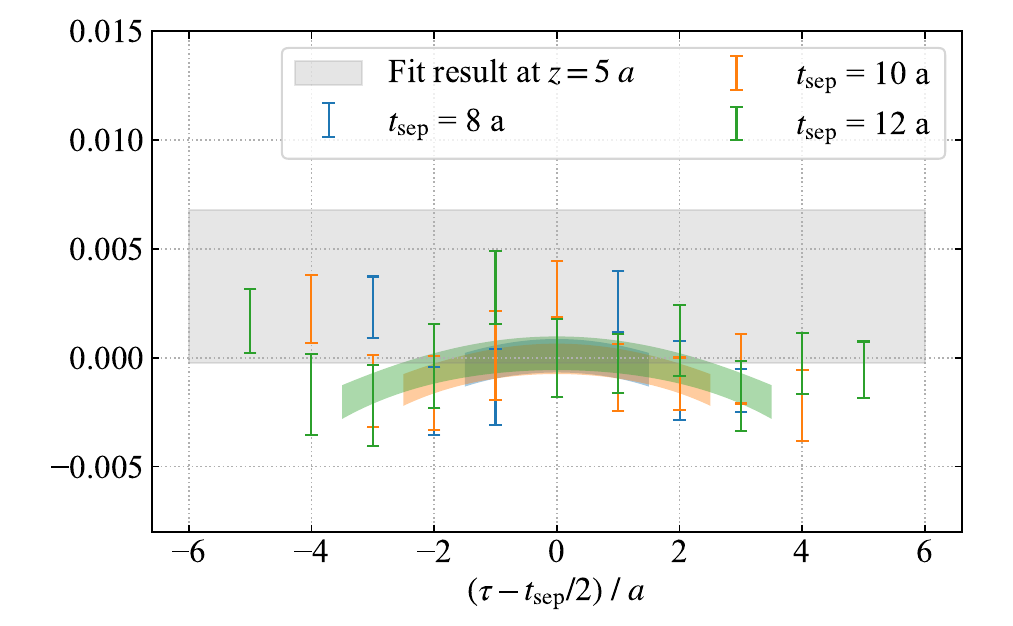}
\includegraphics[width=.4\textwidth]{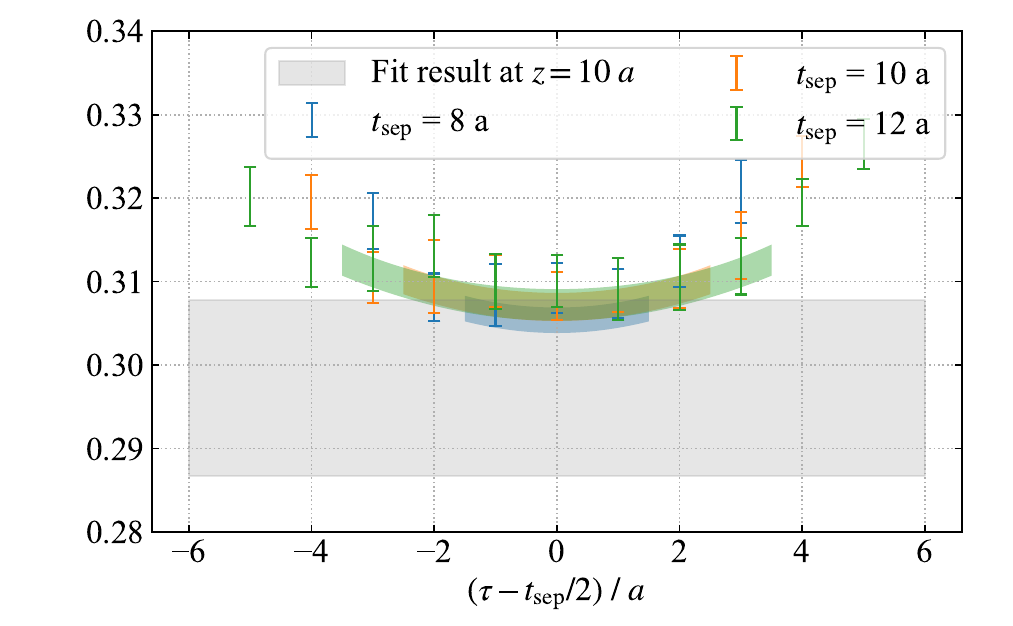}
\includegraphics[width=.4\textwidth]{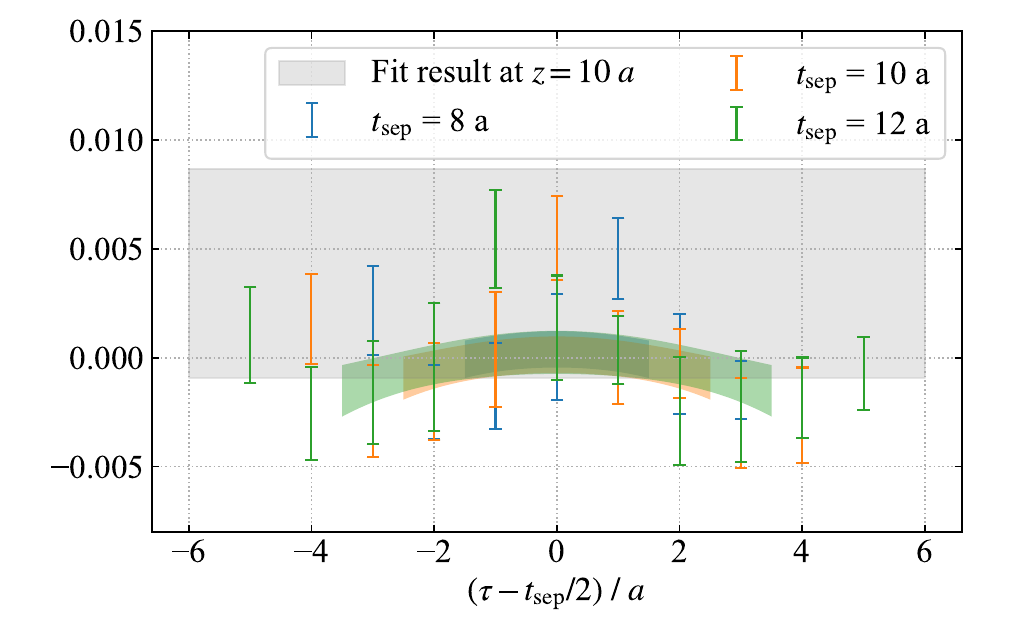}
\caption{Ratio fit of the pion non-local quark correlator using Smallest f. Left column is the real part of the matrix elements and right column is imaginary part of the matrix elements. \label{fig:gsfit_ratio_smallest}}
\end{figure*}

The fit function of the ratio $R(t_{\rm sep}, \tau)$ is
\begin{align}
    R\left(t_{\mathrm{sep}}, \tau \right)=\frac{\sum_{n, m} z_n O_{n m} z_m^{\dagger} \cdot e^{-E_n\left(t_{\text {sep }}- \tau \right)} e^{-E_m \tau}}{\sum_n z_n z_n^{\dagger} \cdot \left( e^{-E_n t_{\text {sep }}} + e^{-E_n (L_t - t_{\rm sep})}\right)} ~,
\end{align}
where $z_n$ are overlap factors and $O_{n m}$ are the matrix elements of operator defined in \Eq{quasi_pdf}, and the summation of $n$ and $m$ are both from $0$ to $1$ according to the \#state in \tb{gsfit}. In order to improve the quality of the ratio fit, a chained fitting method is utilized. This approach involves employing the posterior parameters obtained from two-point fits as the prior parameters for ratio fits. Given that two-point fits exhibit substantially smaller uncertainties, their posterior distribution width is proportionately expanded by a factor of three before being utilized in ratio fits. Most of the fits in all bootstrap samples demonstrate high quality with $\chi^2 / \rm{d.o.f} < 2$. The fit results and data points are shown together in \fig{gsfit_ratio_first} and \fig{gsfit_ratio_smallest}, where a good consistency can be found.

\nocite{*}

\bibliography{main}

\end{document}